\documentclass[lettersize,journal]{IEEEtran}

\usepackage{array}
\usepackage{url}
\usepackage{graphicx}
\usepackage{cite}
\usepackage[colorlinks = true, linkcolor=red, citecolor=red,urlcolor=black]{hyperref}
\usepackage{balance}
\usepackage{soul}
\providecommand{\tabularnewline}{\\}

\graphicspath{{./figures/}}

\begin{document}
	
\title{Five Common Misconceptions About \\Privacy-Preserving Internet of Things}

\author{Mohammad~Abu~Alsheikh,~\IEEEmembership{Senior Member,~IEEE}
\thanks{This work was supported by the Australian Research Council (ARC) under Grant DE200100863. The data collected in this project has been approved by the Human Research Ethics Committee at the University of Canberra under the application ``4522 - Privacy Coupling: When Your Personal Devices Betray You''.}
}

\maketitle

\begin{abstract}

Billions of devices in the Internet of Things (IoT) collect sensitive data about people, creating data privacy risks and breach vulnerabilities. Accordingly, data privacy preservation is vital for sustaining the proliferation of IoT services. In particular, privacy-preserving IoT connects devices embedded with sensors and maintains the data privacy of people. However, common misconceptions exist among IoT researchers, service providers, and users about privacy-preserving IoT. 

This article refutes five common misconceptions about privacy-preserving IoT concerning data sensing and innovation, regulations, and privacy safeguards. For example, IoT users have a common misconception that no data collection is permitted in data privacy regulations. On the other hand, IoT service providers often think data privacy impedes IoT sensing and innovation. Addressing these misconceptions is essential for making progress in privacy-preserving IoT. This article refutes such common misconceptions using real-world experiments and online survey research. First, the experiments indicate that data privacy should not be perceived as an impediment in IoT but as an opportunity to increase customer retention and trust. Second,  privacy-preserving IoT is not exclusively a regulatory problem but also a functional necessity that must be incorporated in the early stages of any IoT design. Third, people do not trust services that lack sufficient privacy measures. Fourth, conventional data security principles do not guarantee data privacy protection, and data privacy can be exposed even if data is securely stored. Fifth, IoT decentralization does not attain absolute privacy preservation.
\end{abstract}


\section*{\textbf{Introduction}}\label{sec:intro}

\IEEEPARstart{R}{ecent} years have witnessed a proliferation of Internet of Things (IoT) services in home automation, retail, telehealth, manufacturing, autonomous vehicles, and precision agriculture. Cisco Systems estimates that 29.3 billion networked devices, 5.7 billion mobile subscribers with an average cellular speed of 43.9 Mbps, and 1.4 billion 5G-enabled devices will exist in 2023~\cite{cisco2020cisco}. This proliferation in ubiquitous computing and high-speed mobile networks enables service providers to collect valuable IoT datasets. In particular, IoT data is fundamental for creating new IoT services tailored to people's needs.

IoT data can contain sensitive data about people, creating genuine data privacy concerns for users. For example, more than 1,272 major data breaches happened in 2019, exposing 163 million records (or 128,171 exposed records per data breach)~\cite{cisco2020cisco}. Data breaches severely impact victims, including financial loss, psychological trauma, criminal impersonation, and online fraud. Accordingly, privacy-preserving IoT applies strict measures to maintain people's data privacy. Specifically, privacy-preserving IoT is a fully-operational IoT network that meets the relevant data privacy regulations, such as the General Data Protection Regulation (GDPR)~\cite{gdpr2016general} and the California Consumer Privacy Act (CCPA)~\cite{ccpa2018california}.

Nonetheless, privacy-preserving IoT has a few commonly misunderstood aspects, as IoT is a technology for pervasive sensing about people and their surroundings. Specifically, common misconceptions exist among IoT service providers and users about the possibility of maintaining data privacy in pervasive IoT sensing. This article refutes common misconceptions, which are organized into three categories—misconceptions about IoT innovation, data regulations, and privacy safeguards. First, IoT is widely motivated as an efficient and cheap method for data collection about people and their surroundings. Thus, data privacy may be wrongly perceived as an impediment to IoT innovation and pervasive sensing. Second, privacy-preserving IoT is often recognized as an exclusive regulatory problem while ignoring its financial benefits and technical requirements. Third, misconceptions exist about the expected benefits of IoT privacy safeguards and decentralization.

The author identifies five common misconceptions about privacy-preserving IoT that appear widely in the literature and recurred in discussions with students, researchers, IoT users, and industry professionals. This article refutes the misconceptions and provides corrections by applying quantitative research methodologies of experimental analysis on real-world datasets and survey research with statistical analysis. First, data privacy does not impede IoT innovation, and a trade-off exists between service accuracy and privacy budget (Misconception~1). Second, data privacy is not exclusively a regulatory problem (Misconception~2). Third, users distrust services that do not make sufficient efforts to protect users' privacy (Misconception~3). Fourth, basic data security principles, i.e.,~the confidentiality, integrity, and availability (CIA) conditions, do not guarantee privacy preservation (Misconception~4). For example, privacy attacks can be exploited to estimate users' faces in exposed facial recognition systems even when the original data is securely stored. Fifth, decentralized IoT (DeIoT) does not provide absolute privacy preservation (Misconception~5).

The rest of this article is organized as follows. First, an overview of privacy-preserving IoT and its main entities is presented. Then, misconceptions about IoT data sensing are discussed. After that, misconceptions about data privacy regulations in privacy-preserving IoT are debated. Common misconceptions about privacy safeguard tools are also debunked. Finally, critical questions for future research directions are discussed, and the article is concluded.

\section*{\textbf{Privacy-preserving IoT: Overview}}\label{sec:sec_1}
\begin{figure*}
	\begin{centering}
		\includegraphics[width=0.95\textwidth,trim=1cm 0.75cm 1cm 1cm]{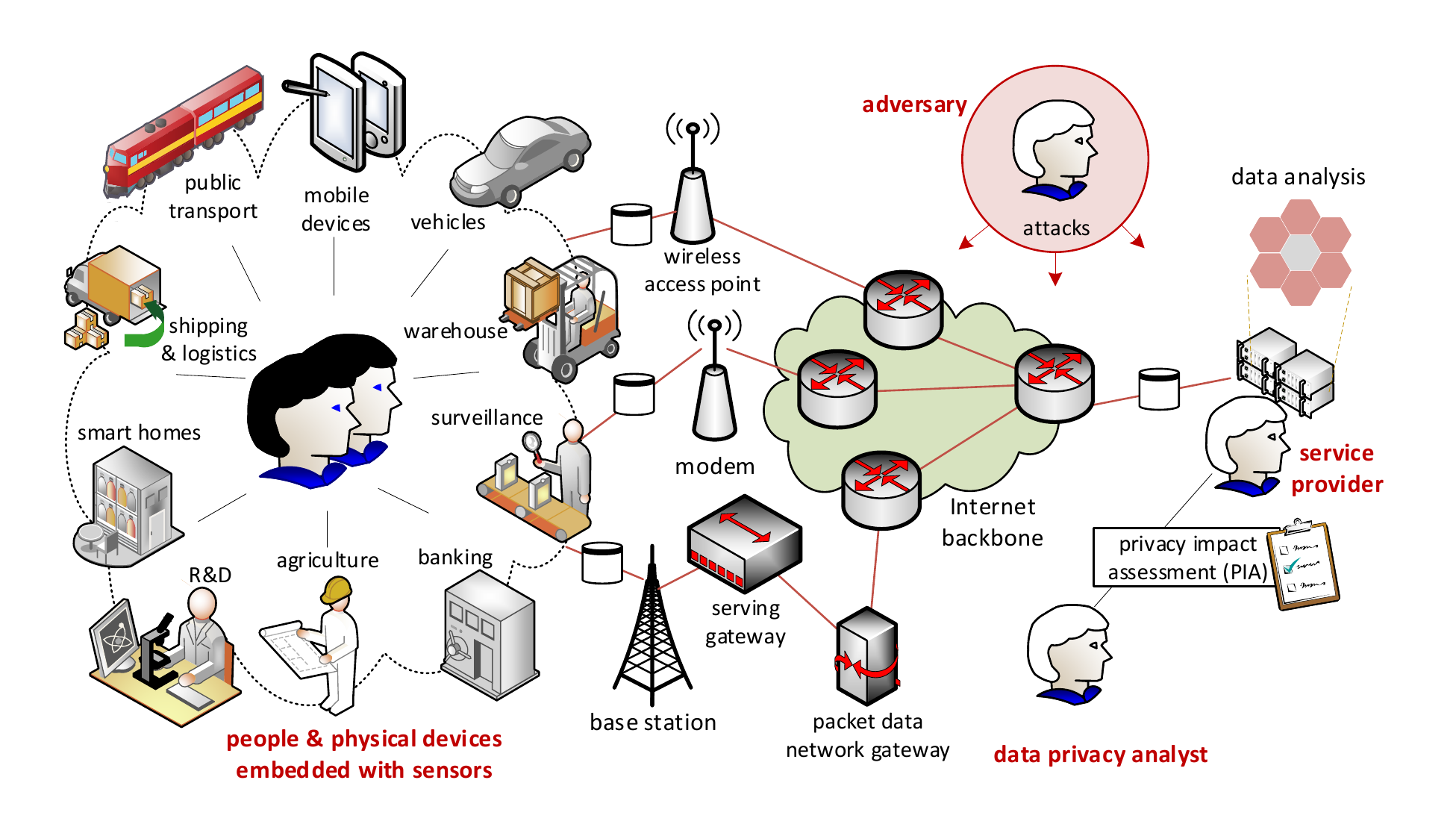}
		\par\end{centering}
	
	\caption{Main entities of privacy-preserving IoT.\label{fig:system_model}}
\end{figure*}

Connecting to IoT services is indispensable and makes people's life more convenient. IoT services are characterized by their ability to interconnect people and physical objects embedded with sensors, actuators, and network connectivity. Real-world examples of IoT services include CropX (www.cropx.com) for real-time crop monitoring in precision agriculture, HealthMap (www.healthmap.org) for crowdsensing healthcare and disease outbreak monitoring, and Fitbits (www.fitbit.com) for fitness tracking.

\subsection*{\textbf{Privacy-preserving IoT}}
\textit{Privacy-preserving IoT refers to any IoT service, i.e., any network of objects embedded with sensors and connection links, that functions while maintaining the privacy rights of users.} Figure~\ref{fig:system_model} shows the four main entities of privacy-preserving IoT.

\begin{itemize}
	\item \textit{People (service users)}: People hold the ownership of their data in privacy-preserving IoT. Each user will possess 3.6 devices in 2023~\cite{cisco2020cisco}. Accordingly, IoT is a cheap and efficient method for large-scale data sensing about people and their surroundings.
	
	\item \textit{Service provider and business stakeholders}: Service providers transmit IoT data to backend servers through high-speed networks. Service providers apply ambient intelligence, including data analysis and machine learning, to create new IoT services and attain personalized user experiences. Business stakeholders apply revenue generation strategies and offer IoT services to interested customers subject to a subscription fee.
	
	\item \textit{Adversary}: An adversary is any third-party entity that initiates privacy attacks to partially or fully attain user data. Many data privacy attacks exist and target all segments of the IoT network, including man-in-the-middle, data poising, membership, and model inversion attacks.
	
	\item \textit{Data privacy analyst (regulatory officer)}: A data privacy analyst oversees the compliance of service providers with the relevant data privacy regulations. A data privacy analyst produces the privacy impact assessment (PIA), which specifies all data flows in a privacy-preserving IoT network and their corresponding data privacy risks. Accordingly, data privacy safeguards, such as data anonymization, perturbation, tokenization, and encryption,  are implemented to mitigate the identified privacy risks. 

\end{itemize}

\subsection*{\textbf{Data privacy rights}}

Recent years have witnessed the introduction of strict data privacy regulations that govern data operations in IoT services. For example, the General Data Protection Regulation (GDPR)~\cite{gdpr2016general} is a European Union law that defines eight data privacy rights of people---the right to be informed of all data operations, the right to review and access copies of personal data, the right to rectify incorrect data, the right to object data processing, the right to restrict data processing, the right of data portability to third parties, the right to be forgotten if personal data is no longer needed for the original purpose, and the right not to be a subject of automation and profiling. Likewise, the California Consumer Privacy Act (CCPA)~\cite{ccpa2018california} defines equivalent user rights for residents of California.

\section*{Misconceptions about IoT data sensing and innovations}\label{sec:sec_2}

Data privacy is widely deemed a fundamental human right. At the same time, IoT is motivated as an affordable and scalable technology for data sensing about people, their surroundings, and everyday activities. This section debunks the misconception that pervasive IoT sensing and data privacy cannot co-exist.

\subsection*{Misconception~1: Data privacy impedes IoT innovation and implies that IoT data cannot be collected}

IoT data is collected to create new IoT services and tailor existing ones for user personalization through ambient intelligence. Therefore, IoT service providers and stakeholders commonly perceive privacy-preserving IoT as an impediment to innovation as data sensing about people is heavily regulated. On the other hand, some service users assume that service providers should not collect any data. These misconceptions arise from an inaccurate understanding of the concept of data privacy.

\textit{Correction}: The rights and responsibilities of each entity in privacy-preserving IoT are well-depicted. Specifically, privacy-preserving IoT is mainly about providing users with control over their data while promoting safeguarded IoT sensing and innovation. Therefore, the service provider must incorporate various privacy safeguards, including explicit consent, rectification forms, and meeting the differential privacy measurements when serving user requests. Privacy-preserving IoT is generally devised to meet the differential privacy requirements~\cite{dwork2014algorithmic} by adding noise to the input data, the parameters of ambient intelligence models, or the output results. Dwork and Roth~\cite{dwork2014algorithmic} describe differential privacy as a guarantee provided by a service provider to users that they will not be affected by sharing their data, regardless of the availability of other information sources or personal data about them, i.e.,~the privacy guarantee is satisfied even when prior knowledge is available about a user. Therefore, this article chooses differential privacy as a state-of-the-art measure for privacy preservation and data leakage in IoT. Moreover, service providers generally apply cryptographic algorithms, e.g.,~tokenization, homomorphic encryption, and secure multi-party computation, to hide sensitive data during computation and data analysis.

\begin{figure}
	\begin{centering}
		\includegraphics[width=0.95\columnwidth,trim=1.2cm 0.75cm 1.5cm 1cm]{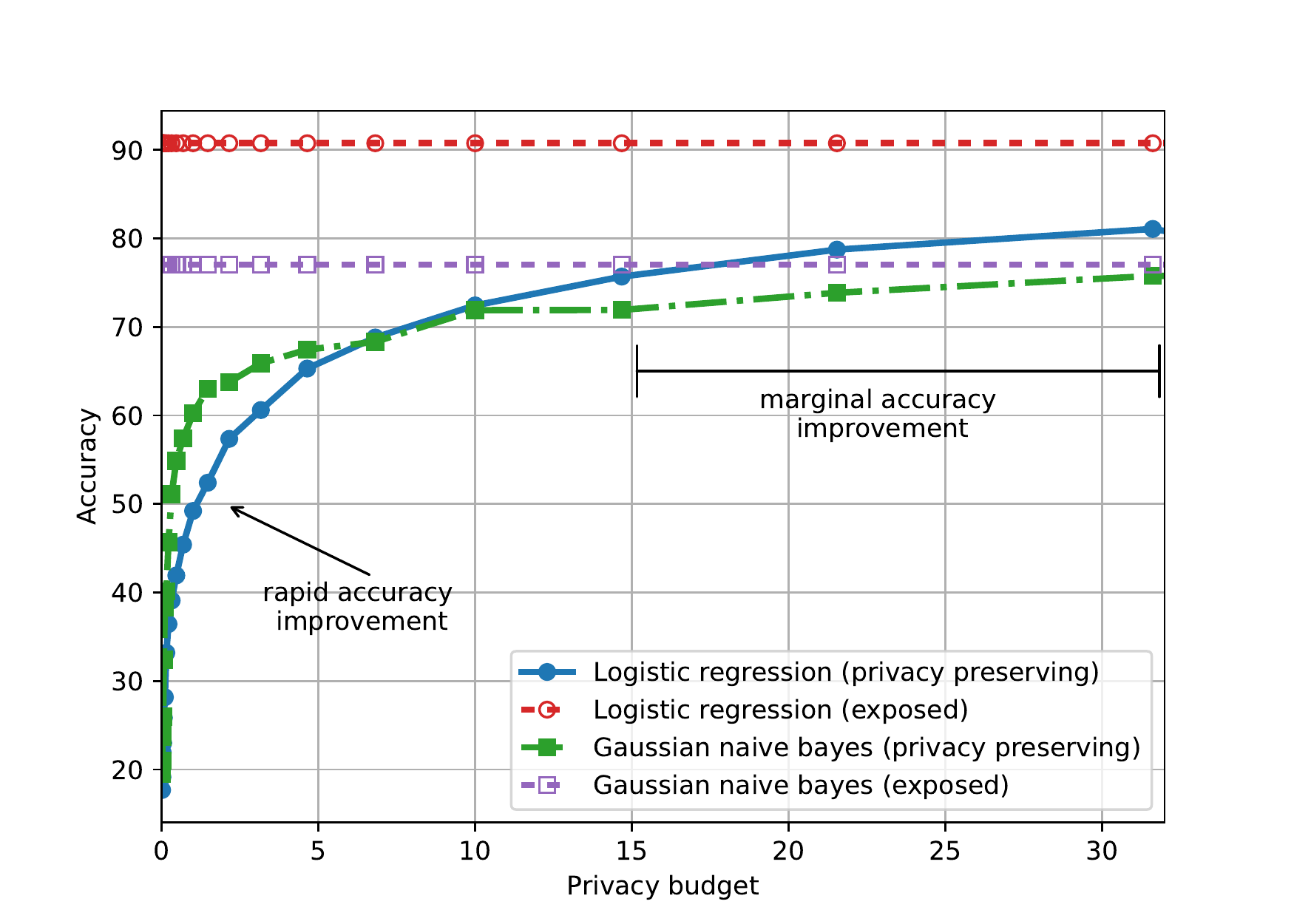}
		\par\end{centering}
	
	\caption{Accuracy of privacy-preserving and exposed IoT services.\label{fig:visualize_privacy_accuracy}}
\end{figure}

Data privacy is not absolute in privacy-preserving IoT. Instead, a trade-off exists between service accuracy and privacy preservation. Figure~\ref{fig:visualize_privacy_accuracy}  shows the accuracy performance of privacy-preserving and exposed IoT services for human activity recognition. Privacy-preserving and exposed services are created using logistic regression and Gaussian naive Bayes trained on a real-world activity prediction dataset~\cite{kwapisz2011activity}. The real-world dataset contains 1,098,207 accelerometer data points of 36 subjects performing everyday activities, including walking, jogging, climbing stairs, sitting, standing, and lying down. Indeed, accelerometer sensors are widely utilized in IoT gadgets and wearables for detecting users' activity in fitness and e-health applications. The \textit{privacy budget} in differential privacy is defined as the probability of accidental data leakage by an adversary. A small privacy budget requires adding significant noise to the machine learning parameters, resulting in high privacy preservation. Cross-validation of 5 folds was used in each experiment, and each experiment was repeated 20 times.

Several important results can be noted from Figure~\ref{fig:visualize_privacy_accuracy}. First, when the privacy budget increases, the service accuracy will increase. This relationship is expected because satisfying the tight privacy budgets needs adding more noise; thus, the service accuracy will decrease. Second, when the privacy budget is small, i.e., less than 5, significant accuracy improvement can be achieved for small increases in the privacy budget requirements. Third, there is a marginal gain in the service accuracy for increasing the privacy budget at high values. Fourth, different algorithms may have different accuracy ranks when changing the privacy budget. For example, Gaussian naive Bayes has higher accuracy than logistic regression when the privacy budget is less than 6.8. Then, logistic regression reports higher accuracy values when the privacy budget exceeds 6.8. Finally, the exposed services retain higher accuracy values than the privacy-preserving ones, but that accuracy gain comes at the cost of risking users' privacy.

\section*{\textbf{Misconceptions about data privacy regulations}}\label{sec:sec_3}

IoT standards, e.g.,~IEEE 2413-2019~\cite{ieee2020std}, include data privacy as a functional requirement of IoT architectures. This section discusses two misconceptions about data privacy regulations and IoT. The first misconception emphasizes that privacy-preserving IoT is an exclusive regulatory problem. The second misconception undervalues the benefits of privacy-preserving IoT in assembling trust bridges with users and improving customer retention.

\subsection*{\textbf{Misconception~2: Privacy-preserving IoT is exclusively a regulatory problem}}

The privacy paradox is a widely used concept in the literature to describe the discrepancy between how people insist on the importance of their privacy and how they compromise their privacy in reality. For example, many users still provide their names and emails in marketing campaigns to receive discounts or free product samples. Accordingly, data privacy has been portrayed as an exclusive regulatory problem, i.e., people are wrongly perceived as incompetent in protecting their privacy. Therefore, Solove \cite{solove2021myth} argues for regulating service architectures and against privacy self-management, describing it as a complex task for users.

\textit{Correction}: Privacy-preserving IoT is not an exclusive regulatory problem. Two main issues regarding restricting user-level data control can be underlined. First, data privacy is individual-level ownership rather than a societal right. Accordingly, people should be able to provide consent to service providers for data collection and selling. Existing privacy regulation, such as the California Consumer Privacy Act (CCPA)~\cite{ccpa2018california}, underlines that users may be offered discounts and financial incentives for data collection. This arrangement provides flexibility to both users and service providers. Second, a single government body cannot check the compliance of every service provider with the privacy regulations. The centralized privacy authority creates an unnecessary bottleneck in privacy-preserving IoT. Third, many sensing technologies exist in IoT systems, and it would be unattainable for a single entity to assess all possible privacy risks.

Data privacy must be incorporated in the early design cycle of IoT. Internal and external policies must be drafted early in the design cycle. In addition, a data privacy analyst must be recruited to devise privacy impact assessments and breach response plans to comply with the data privacy regulations.

\subsection*{\textbf{Misconception~3: Privacy-preserving IoT is exclusively required to comply with data privacy regulations}}

A common misconception among service providers is perceiving data privacy in IoT as an obligation that does not retain direct financial benefits. Therefore, service providers adhere to the data privacy regulations as a compliance action, and IoT data privacy is not perceived as a functional requirement. Notably, this misconception underestimates the significance of privacy preservation in customer satisfaction, retention, and trust.

\textit{Correction}: Privacy-preserving IoT has many benefits for building trust bridges with users; hence, it boosts user retention and satisfaction. An online survey study was conducted to understand how people perceive their data privacy in exposed systems. The survey was created using the Qualtrics platform (\url{www.qualtrics.com}). The survey responses were collected from 200 participants recruited using Amazon Mechanical Turk (\url{www.mturk.com}) for crowdsourcing task assignments.

\begin{table}
	
	\caption{Users take various actions if a company does not make sufficient efforts to protect their online data privacy.}\label{tab:privacy_actions}
	\setlength\extrarowheight{4pt}
	\begin{centering}
		\begin{tabular}{|>{\raggedright}p{0.65\columnwidth}|>{\raggedright}p{0.25\columnwidth}|}
			\hline 
			\textbf{user action \\(a user can select multiple actions)} & \textbf{percentage \& \\95\% confidence interval (\%)}\tabularnewline
			\hline 
			\hline 
			Stop using the company's services & 56.2 $[49.3, 62.9]$\tabularnewline
			\hline 
			Close service accounts & 67.2 $[60.4,73.3]$\tabularnewline
			\hline 
			Request deleting data & 66.7 $[59.9,72.8]$\tabularnewline
			\hline 
			Report the company to cybersecurity services and government agencies,
			such as the National Cybersecurity Center & 41.3 $[34.7,48.2]$\tabularnewline
			\hline 
			Share the information with family and friends so they can avoid using
			the company's services & 37.8 $[31.4,44.7]$\tabularnewline
			\hline 
			Unsubscribe from the company's email list & 48.3 $[41.4,55.1]$\tabularnewline
			\hline 
			None & 5.0 $[2.7,8.9]$\tabularnewline
			\hline 
		\end{tabular}
	\par\end{centering}
	
\end{table}

\begin{itemize}
	\item \textit{People will not use exposed services}: Table~\ref{tab:privacy_actions} shows the response percents and 95\% confidence intervals of the various actions that users would take if a company does not make sufficient efforts to protect their data privacy. A 95\% confidence interval indicates the range of likely values of the responses for which the survey results are accurate with a 95\% confidence level, i.e., 19 times out of 20 repetitions. The respondents indicated they would take all possible measures to protect their privacy. For example, 56.2\% suggested that they would stop using the company's services, and 67.2\% said they would close the service accounts. Only 5.0\% of the respondents would not take any action. The 95\% confidence intervals show that the survey results are accurate and represent the general population with high confidence. For example, the probability of a user to "stop using the company's services" in the confidence interval [49.3\% to 62.9\%]  is accurate under the confidence level of 95\%. The survey responses show that exposed services will lose many users due to the lack of proper privacy measures.
	
	\item \textit{92.5\% of people are genuinely concerned about their data privacy and how service providers use their online data}: 42.5\% and 50.0\% of the respondents reported that they are ``strongly concerned'' and ``somewhat concerned'' about their data privacy and how service providers use their online data. The ``strongly concerned'' and ``somewhat concerned'' responses have 95\% confidence intervals of [35.9\% to 49.4\%] and [43.1\% to 56.9\%], respectively. Only 2.5\% of the respondents are unconcerned about their privacy. 5.0\% of the respondents are neither concerned nor unconcerned about their data privacy.
	
	\item \textit{73\% of people do not trust companies that do not make sufficient efforts to protect their data privacy}: The survey results show that data privacy is essential for gaining users' trust. Notably, 29.5\% and 43.5\% of the respondents indicated that they ``strongly distrust'' and ``somewhat distrust'' companies that do not make sufficient efforts to protect their data privacy with 95\% confidence intervals of [23.6\% to 36.2\%] and [36.8\% to 50.4\%], respectively. Only 8.5\% of the respondents trust exposed companies. In addition, 18.5\% of the respondents do not have strong opinions.
	
\end{itemize}

The survey research's results indicate the importance of data privacy in improving user retention and overall satisfaction. Thus, data privacy should not be perceived as a compliance problem but rather as a business opportunity with financial yields.

\section*{\textbf{Misconceptions about privacy safeguards}}\label{sec:sec_3}

There has been significant progress in data security technologies and privacy safeguards. However, a common misunderstanding exists regarding the relationship between data security and data privacy in IoT. In particular, fulfilling the security principles, i.e.,~the confidentiality, integrity, and availability (CIA) conditions, does not mean data privacy is protected. This section debunks two myths about technical solutions for meeting the data privacy requirements in privacy-preserving IoT. First, this section shows that data privacy may not be met even if IoT data is securely stored. Then, it shows that IoT decentralization does not guarantee absolute data privacy for users.

\subsection*{\textbf{Misconception~4: Data privacy is fully preserved if IoT data is securely stored}}

A widespread fallacy, even among cybersecurity practitioners, is claiming data privacy preservation by applying data security measures, such as network security, access control, backups, authorization, firewalls, and intrusion detectors. In particular, such security methods are implemented to adhere to the confidentiality, integrity, and availability (CIA) principles. Confidentiality protocols, e.g., access control and authorization, aim to protect the data from unauthorized disclosure. Integrity,  e.g., digital signatures and logging, aims to maintain the accuracy and completeness of data. Finally, availability, e.g., backups and firewalls, aims to promptly supply resource access to users when requested.

\begin{figure}
	\begin{centering}
		\includegraphics[width=0.9\columnwidth,trim=0.5cm 0.75cm 0.5cm 1cm]{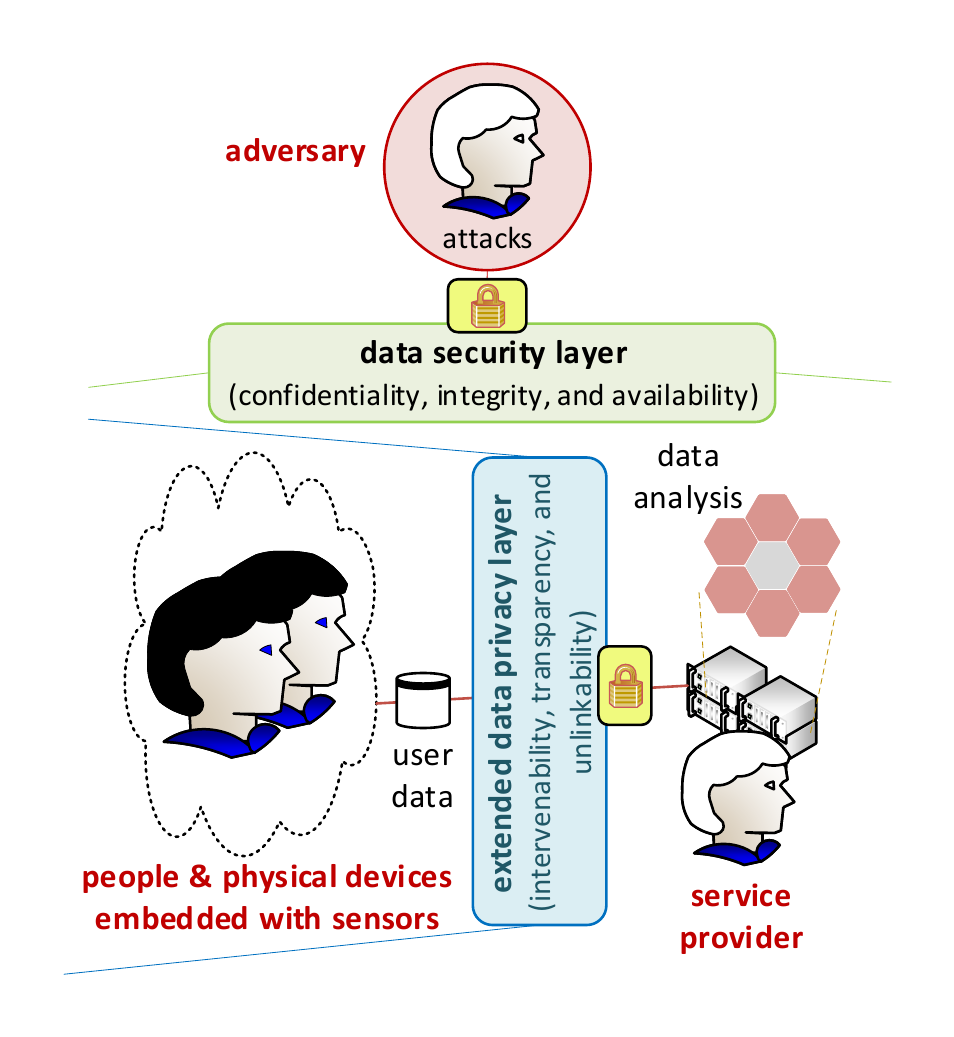}
		\par\end{centering}
	
	\caption{Data privacy extends the data security conditions, providing users with control over their data and preventing data violations and misuse.\label{fig:data_privacy_security}}
\end{figure}

\textit{Correction}: Data security, defined in the confidentiality, integrity, and availability (CIA) triad, does not guarantee users' data privacy. In particular, data security protects users from unauthorized data access or modification. On the other hand, data privacy protects users from violations and misuse, including how service providers use and process user data. Data privacy is a superset of data security and requires stricter conditions to comply with the privacy laws on how user data is collected, transmitted, stored, and processed, e.g.,~the data privacy rights of users as depicted in the General Data Protection Regulation (GDPR)~\cite{gdpr2016general} and the California Consumer Privacy Act (CCPA)~\cite{ccpa2018california}. For example, the confidentiality, integrity, and availability (CIA) conditions do not cover the rights to be informed or forgotten, which are fundamental privacy rights of users. Therefore, Hansen~\textit{et al.}~\cite{hansen2015protection} extend the confidentiality, integrity, and availability (CIA) triad to support data privacy by including intervenability, transparency, and unlinkability conditions. Intervenability enables the users to intervene and provoke their data rights, such as consent withdrawal and data rectification and erasure. Then, transparency ensures that users can understand and verify all data operations with reasonable effort, which enables users to provide informed consent for data operations. Finally, the unlinkability condition controls linking user data with information from other sources, preventing the risk of user re-identification and automated profiling. Figure~\ref{fig:data_privacy_security} depicts how data privacy extends data security, providing users with control over their data and preventing data violations and misuse.

\begin{figure}
	\begin{centering}
		\includegraphics[width=0.98\columnwidth,trim=0cm 0cm 0.5cm 1cm]{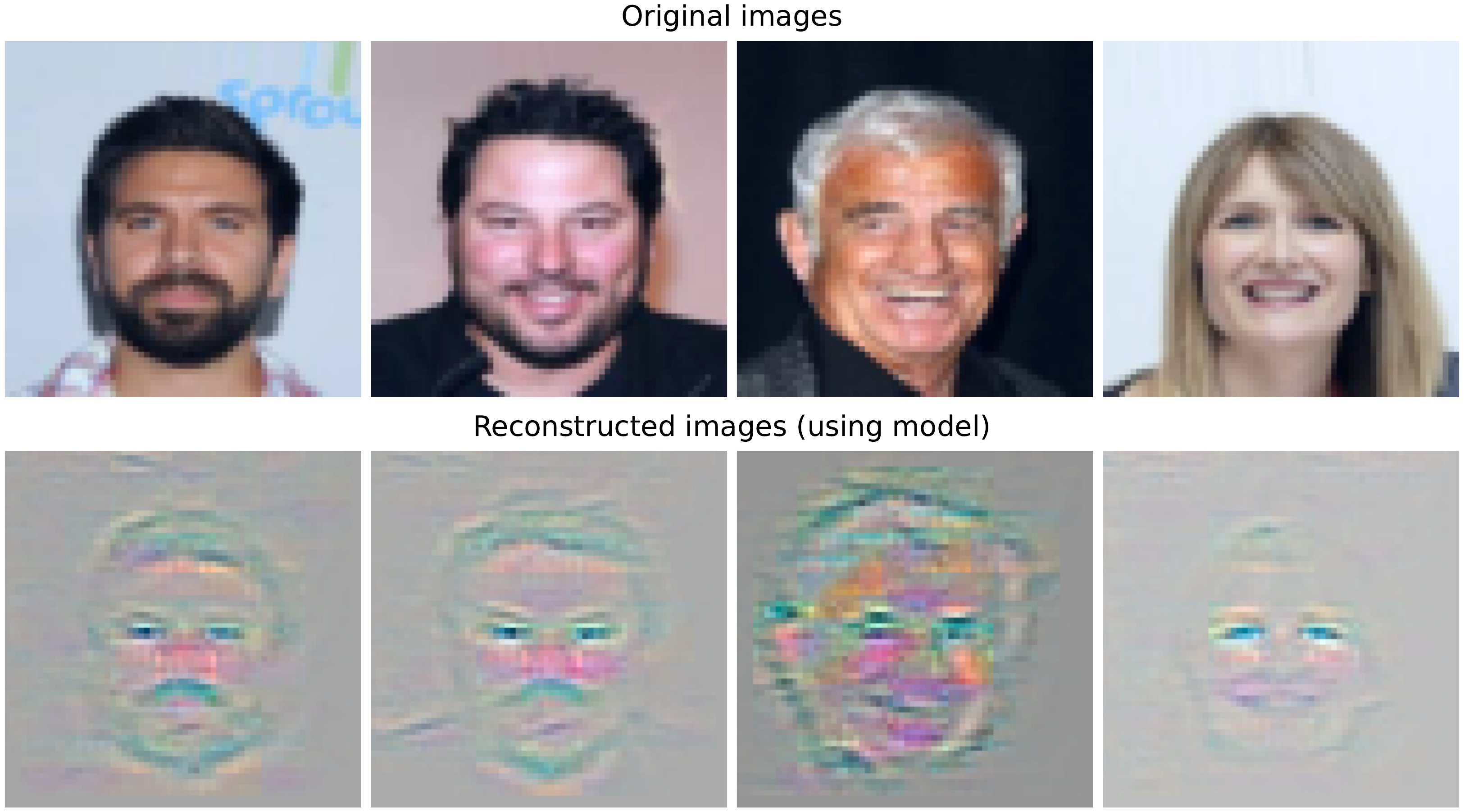}
		\par\end{centering}
	
	\caption{Privacy attacks on an exposed IoT service that uses face recognition.\label{fig:visualize_privacy_attack}}
\end{figure}

Data privacy may not be met even when original data is securely stored. This can be demonstrated using a facial recognition system that is widely used for easy sign-in to IoT services. Figure~\ref{fig:visualize_privacy_attack} depicts how an adversary can reconstruct people's faces in exposed IoT facial recognition systems. The facial recognition service is built by training a deep learning model on the CelebFaces dataset~\cite{liu2015deep}, which contains 202,599 images of 10,177 identities. Even though the original face images are securely kept, the adversary can reconstruct an accurate estimation of people's faces using the deep learning model, i.e., the original training images are not used in producing the reconstructed images. Such a privacy attack is an example of model inversion attacks~\cite{he2020towards} that produce sensitive data using outputs of a model. Figure~\ref{fig:visualize_privacy_attack} demonstrates how data privacy is not fully preserved even when the confidentiality, integrity, and availability (CIA) conditions are satisfied. The attack arises in many real-world IoT systems that ship trained deep learning models with IoT consumer products, e.g., for running face recognition with edge computing at IoT devices. Accordingly, an adversary can effortlessly obtain a copy of the trained deep learning model and apply model inversion attacks. Therefore, service providers should utilize privacy-preserving learning that adds reasonable noise to the modeling parameters during model training according to the differential privacy conditions~\cite{dwork2014algorithmic}. Moreover, learning from encrypted data using homomorphic encryption and secure multi-party computation must be utilized to thwart model inversion attacks and the reconstruction of user data.

\subsection*{\textbf{Misconception~5: Decentralized IoT (DeIoT) solves the privacy problem and provides absolute data privacy preservation}}

Decentralized IoT (DeIoT) is an emerging user-centered ecosystem that distributes IoT control functions and delegates operations to users without including a central authority. Edge computing, blockchain ledgers, and federated learning are the most promising technologies for DeIoT. For example, smart contracts and blockchain ledgers provide decentralized digital identities that are shared with all participating devices~\cite{nguyen2020trusted}. In addition, federated learning and edge computing can optimize a master ambient intelligence model without sharing users' original data with a central server~\cite{malandrino2021federated}. DeIoT is widely suggested as a method for improving data privacy, security, transparency, and scalability using token-based operations and decentralization. As a result, DeIoT allows people to gain additional control over their data collection and processing.

\begin{figure}
	\begin{centering}
		\includegraphics[width=0.9\columnwidth,trim=2cm 0.75cm 2cm 1cm]{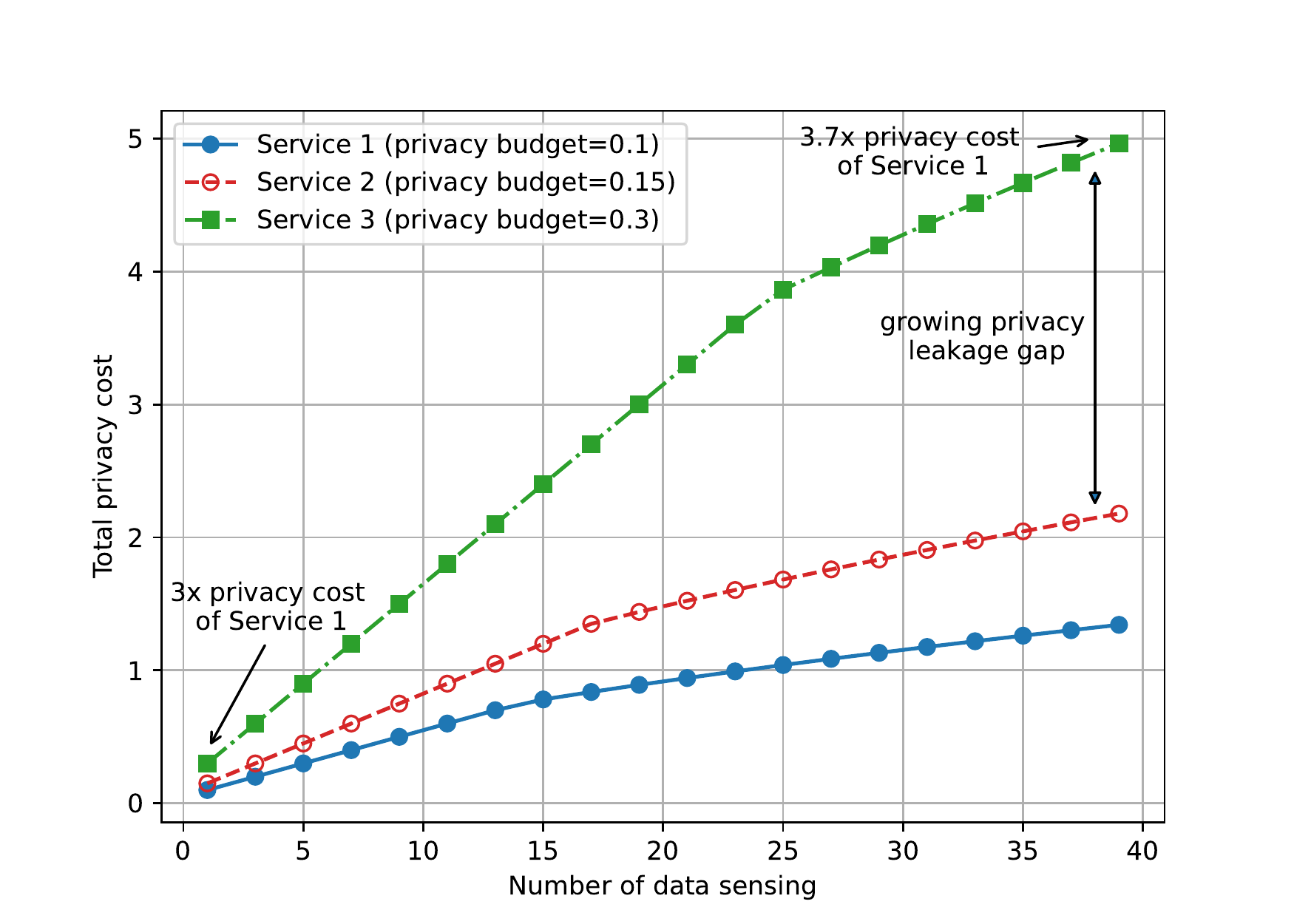}
		\par\end{centering}
	
	\caption{Total privacy cost of repeated data sensing at various privacy budgets.\label{fig:visualize_privacy_sensing}}
\end{figure}

\textit{Correction}: Unfortunately, DeIoT does not provide absolute data privacy preservation. First, DeIoT is challenging to regulate, given the lack of a central moderation entity. Second, adversaries can impersonate legitimate users, and harmful content can be injected into DeIoT networks. Third, DeIoT introduces new privacy and security challenges. For example, blockchains and federated learning are two prominent DeIoT-enabling technologies that raise new privacy risks in IoT services. Blockchain uses asymmetric cryptography, i.e.,~pairs of private and public keys, to securely store transactions on a public ledger that all users can view. Therefore, user identities in DeIoT can be impersonated if the user's private key, which is unique for each user, is stolen or hacked by an adversary. Then, blockchain-based DeIoT guarantees immutability~\cite{nguyen2020trusted}, indicating that IoT data cannot be altered or deleted once a blockchain record is verified. Nonetheless, the immutability of DeIoT violates the user rights in data erasure and rectification as depicted in the General Data Protection Regulation (GDPR)~\cite{gdpr2016general} and the California Consumer Privacy Act (CCPA)~\cite{ccpa2018california}. Also, adversaries can participate in federated learning to access the final trained model, creating privacy risks that can be exploited using model inversion attacks~\cite{he2020towards}.

When the privacy budget equals zero, absolute data privacy is achieved, and differential privacy guarantees that an adversary cannot identify data about individual users. As discussed above, DeIoT does not fulfill the requirements of absolute data privacy. Assume that three services (Services~1-3) are built using blockchain ledgers. The privacy budget of a single data sensing is set at 0.1, 0.15, and 0.3 in Services~1-3, respectively. Figure~\ref{fig:visualize_privacy_sensing} shows the total privacy costs of accessing Services~1-3 at different privacy budgets. The total privacy cost indicates the cumulative privacy budget of differential privacy for sequential sensing events in IoT, i.e.,~overall privacy leakage as a function of the number of data sensing events. The total privacy cost is calculated using the composition theorem of privacy-preserving mechanisms~\cite{dwork2014algorithmic}. Two substantial results can be underlined. First, the total privacy cost increases over repeated sensing in the three services. Therefore, although the privacy budget is small for a single sensing event, data privacy preservation in DeIoT is not absolute. Second, the difference in the privacy cost of users magnifies over time. For example, Service 3 has 3.7x the privacy cost of Service 1 after 38 data sensing attempts, even though the privacy budget of a single sensing attempt in Service 3 is 3x that of Service~1.

\section*{\textbf{Critical questions for future research}}\label{sec:sec_5}

This section presents critical questions for future research in privacy-preserving IoT.

\subsection*{\textbf{Data privacy and criminal justice}}

A widespread argument for supporting dataveillance, i.e.,~monitoring and profiling people's data, is for criminal justice, law enforcement, and fraud prevention. For example, all United States privacy laws exclude law enforcement from information disclosure even though the Fourth Amendment protects individuals from unreasonable searches by the government without a reasonable basis~\cite{murphy2013politics}. Furthermore, privacy-preserving IoT might be seen as a tool for boosting online incivility and unfriendly behavior. Such arguments for dataveillance undervalue the benefits of data privacy. First, data privacy is a powerful mechanism for preventing cybercrimes and online incivility. In particular, when personal data is privately reserved, people will be immune to adversaries and disruptive behaviors. Second, data privacy is a crucial enabler for the freedom of speech and civil liberty. Third, organized crime has the resources and motives to create custom encrypted communication channels; thus, exposed services will have the most impact on regular users.

Social benefits do not wipe out the personal benefits of data privacy. Several critical questions need further research. What is the proper procedure for requesting data disclosure for criminal justice? How can protected data be accessed for criminal justice without establishing an encryption backdoor? How can people oversee the levels of dataveillance by organizations and governments?

\subsection*{\textbf{User-in-the-loop (UIL) privacy-preserving IoT}}

IoT standards, e.g.,~IEEE 2413-2019~\cite{ieee2020std}, emphasize that user privacy and trust are essential components of any IoT design. However, people, i.e., data owners, are still not well-engaged in their privacy preservation. As discussed in \textit{Misconception~2}, the privacy paradox suggests doubt about the ability of users to protect their privacy. Moreover, users generally cannot verify the privacy measures taken by service providers due to the lack of transparency in the implemented privacy safeguards~\cite{zheng2018user}.

User-in-the-loop (UIL) data privacy engages users in their privacy preservation. UIL is an emerging design concept in IoT that improves performance, e.g., reducing overall energy consumption, by engaging people in fundamental IoT operations beyond traffic consumption and generation~\cite{petrov2018iot}. UIL privacy-preserving IoT has garnered limited scholarly attention. Thus, a few critical questions required further exploration. How can user awareness of data privacy issues be increased? How can service providers provide people with data privacy measurements? How can users be incentivized to contribute to their data protection efforts?


\section*{\textbf{Conclusions}}\label{sec:conclusions}

Privacy-preserving IoT is essential given the substantial data privacy risks and their severe impacts on users. This article presented five common misconceptions about privacy-preserving IoT. Several results were presented using experiments on real-world datasets and survey research. First, a trade-off exists between privacy preservation and service accuracy in privacy-preserving IoT. Second, privacy-preserving IoT is beneficial for user retention and overall satisfaction. Third, users are concerned about their privacy and do not trust services that lack sufficient privacy procedures. Fourth, privacy attacks may occur even when user data is securely stored. Fifth, IoT decentralization does guarantee absolute data privacy. Finally, critical questions were raised for future research in privacy-preserving IoT.


\section*{\textbf{Biographies}}

\balance
\bibliographystyle{IEEEtran}
\bibliography{reference}


\begin{IEEEbiographynophoto}
	{Mohammad Abu Alsheikh}
	[S'14, M'17, SM'22] is an Associate Professor and ARC DECRA Fellow at the University of Canberra, Australia.
\end{IEEEbiographynophoto}
\vfill

\end{document}